\documentclass[pra,preprint,superscriptaddress,floatfix,showpacs]{revtex4-1}

\usepackage{amsmath,amsfonts,amssymb}
\usepackage{graphicx}
\usepackage{tabularx}
\usepackage{bbold}
\def\beq{\begin{equation}}
\def\eeq{\end{equation}}
\def\beqn{\begin{eqnarray}}
\def\eeqn{\end{eqnarray}}

\begin{document}

\title{Many-body effects in the excitation spectrum of weakly-interacting Bose-Einstein condensates in one-dimensional optical lattices}

\author{Raphael Beinke}
\email{raphael.beinke@pci.uni-heidelberg.de}
\affiliation{Theoretische Chemie, Physikalisch-Chemisches Institut, Universit\"at Heidelberg, Im Neuenheimer Feld 229, D-69120 Heidelberg, Germany}

\author{Shachar Klaiman}
\affiliation{Theoretische Chemie, Physikalisch-Chemisches Institut, Universit\"at Heidelberg, Im Neuenheimer Feld 229, D-69120 Heidelberg, Germany}

\author{Lorenz S. Cederbaum}
\affiliation{Theoretische Chemie, Physikalisch-Chemisches Institut, Universit\"at Heidelberg, Im Neuenheimer Feld 229, D-69120 Heidelberg, Germany}

\author{Alexej I. Streltsov}
\affiliation{Theoretische Chemie, Physikalisch-Chemisches Institut, Universit\"at Heidelberg, Im Neuenheimer Feld 229, D-69120 Heidelberg, Germany}

\author{Ofir E. Alon}
\affiliation{Department of Physics, University of Haifa at Oranim, Tivon 36006, Israel}

\date{\today}

\begin{abstract}
In this work, we study many-body excitations of Bose-Einstein condensates (BECs) trapped in periodic one-dimensional optical lattices. In particular, we investigate the impact of quantum depletion onto the structure of the low-energy spectrum and contrast the findings to the mean-field predictions of the Bogoliubov-de Gennes (BdG) equations. Accurate results for the many-body excited states are obtained by applying a linear-response theory atop the MCTDHB (multiconfigurational time-dependent Hartree method for bosons) equations of motion, termed LR-MCTDHB. We demonstrate for condensates in a triple well that even weak ground-state depletion of around $1\%$ leads to visible many-body effects in the low-energy spectrum which deviate substantially from the corresponding BdG spectrum. We further show that these effects also appear in larger systems with more lattice sites and particles, indicating the general necessity of a full many-body treatment. 
\end{abstract}


\maketitle

\section{Introduction}\label{sec:Intro}
The understanding of static and dynamic properties of Bose-Einstein condensates (BECs) in trapped dilute ultracold atomic gases has been a major research objective since the experimental realization of BECs \cite{anderson1995observation,Hulet1995,davis1995bose,RMP1,RMP2}. The standard theoretical approach for these systems is to solve the Gross-Pitaevskii (GP) mean-field equation which assumes all bosons to reside in the same single-particle state \cite{Gross,Pitaevskii,Pitaevskii1,Pethick}. 

Excited states have commonly been accessed by applying linear-response theory atop the GP equation, yielding the Bogoliubov-de Gennes (BdG) equations \cite{Bogoliubov,Pitaevskii1,Pethick,Castin,Gardiner,Castin2}. Several experiments where the Bogoliubov excitations in BECs were studied are reviewed in \cite{Steinhauer}. However, it has been observed for various systems that many-body effects play a crucial role. This concerns ground-state depletion and fragmentation in finite systems \cite{ALS,baderFragmentation,fischerFragmentation} as well as the out-of-equilibrium dynamics of trapped BECs \cite{sakmann2009exact,Streltsov,Peter1,Benchmarks,ALT,tsatsos2014vortex,Beinke_et_al}, even for almost fully-condensed systems where the GP equation was expected to accurately describe the condensate's time-evolution \cite{uncertainty_product}. 

In particular, the dynamics of BECs in one-dimensional (1D) optical lattice potentials have been of high interest over the last two decades \cite{Burger,Bloch,Oberthaler,Schmelcher1}. Some experimental observations of excited states in such systems clearly deviate from the BdG predictions. As an example, the out-of-equilibrium dynamics in a large 1D lattice with a superimposed harmonic confinement shows substantial differences with respect to the observed oscillation frequencies of the superfluid phase and the BdG frequencies \cite{Kasevich1,Kasevich2}. Moreover, the quantum depletion in gaseous BECs exposed to a 1D optical lattice only partly agrees with the BdG theory \cite{Ketterle}. 

In terms of many-body excitations, a contemporary study shows the low-energy spectrum of excited states for four bosons in a triple-well potential, utilizing a number-state expansion of the exact wavefunction \cite{Schmelcher2}. Nonetheless, to our knowledge a general study on the many-body effects in the low-energy spectrum of finite weakly-interacting BECs in one-dimensional lattices is still missing. Our motivation for this work is to exactly fill this gap. 

Analytic results from a recent study show that the single-particle excitation energies of a weakly-interacting BEC in a trap are given exactly by the BdG predictions in the infinite-particle (or mean-field) limit \cite{Seiringer1} (generalizing a previous study on homogeneous systems \cite{Seiringer2}). The energies of excited states where more than one particle is excited out of the condensate are then obtained by multiples and sums of the BdG energies. It remains unclear, however, what are the excitation energies of a BEC in a trap with only a finite amount of particles, far away from the infinite-particle limit. We address this question in this work. Beside discussing general many-body effects in the low-energy spectra, we investigate how far the BdG mean-field energies deviate from the exact many-body results, especially when the system's ground-state is almost fully condensed and a mean-field approach seems to be adequate.

To this end, we discuss the many-body excited states of repulsive BECs in a shallow and deep triple well and present numerically converged results for the energy levels. By comparing the obtained many-body spectra to the corresponding BdG predictions, we demonstrate that already in the limit of weak interaction and weak ground-state depletion of the order of $1\%$, many-body effects occur which can not be explained by the BdG theory.  We emphasize that in our study we are far away from the regime of the superfluid to Mott-insulator transition \cite{Esslinger,Zoller,Greiner}. Furthermore, we show that the same many-body effects also appear in larger lattices where the number of particles and lattice sites is increased, indicating that a many-body description for the excited states in these systems is unavoidable. 

We compute exact values for low-lying many-body excitations by employing linear-response theory atop the multiconfigurational time-dependent Hartree method for bosons (MCTDHB \cite{MCTDHB1,MCTDHB2}), termed LR-MCTDHB, which has been introduced in Refs. \cite{LR-MCTDHB1,LR-MCTDHB2}, successfully benchmarked in Ref. \cite{LR-MCTDHB3}, and recently applied for BECs trapped in harmonic and double-well potentials \cite{Theisen}. In this work, we report on the development and application of our new numerical implementation of LR-MCTDHB, and show how it enables us to find new physics.

The paper is structured as follows. In Sec. \ref{sec:Theo}, we describe the theoretical framework, i.e., the system's Hamiltionian (Sec. \ref{Hamiltonian}) and elaborate on the many-body linear-response method utilized (Sec. \ref{methods}). Results are presented in Sec. \ref{Results and analysis}. We separate the discussion of excited states of BECs in a triple well (Sec. \ref{triple_well}) and in larger systems where  the number of lattice sites and bosons in the trap is increased (Sec. \ref{larger_systems}). A summary as well as concluding remarks are given in Sec. \ref{Conclusions}. Further information on the numerical convergence of our results, a special type of excited states (zero quasi-momentum modes) and a benchmark of LR-MCTDHB to the analytically solvable harmonic interaction model are provided in the Appendixes \ref{Appa}, \ref{AppB} and \ref{AppC}.

\section{Theoretical framework}\label{sec:Theo}

\subsection{Hamiltonian and setup}\label{Hamiltonian}
In this work, we consider condensates of $N$ interacting bosons described by the general Hamiltonian
\begin{equation}\label{Ham1}
	\hat{H}=\sum_{i=1}^N \hat{h}(x_i)+\sum_{i<j=1}^N \hat{W}(|x_i-x_j|)
\end{equation}
with the one-body operator $\hat{h}(x)=-\frac{1}{2}\Delta+\hat{V}(x)$, comprising the kinetic energy and the external single-particle potential $\hat{V}$, and with the two-body interaction potential $\hat{W}$. For the sake of simplicity, we set $\hbar=m=1$ where $\hbar$ is Planck's constant and $m$ is the boson mass. Dimensionless units are obtained by dividing $\hat{H}$ by $\frac{\hbar^2}{d^2m}$ with $d$ being a length scale.  

The external potential represents a one-dimensional lattice given by
\begin{equation}\label{pot}
	\hat{V}(x)=V_0\,\cos^2\left(\frac{\pi}{l}\,x\right)
\end{equation}
where $V_0$ is the lattice depth and $l$ the distance between two adjacent lattice sites. We assume periodic boundary conditions and separate the lattice sites by a distance of $l=1$. $V_0$ can be expressed in terms of dimensionless units of the recoil energy $E_R=\frac{\hbar^2 k_0^2}{2m}$ with the lattice momentum $k_0=\frac{\pi}{l}$.

The interaction between the bosons is described by contact interaction
\begin{equation}
	\hat{W}(|x_i-x_j|)=\lambda_0\, \delta(x_i-x_j) 
\end{equation}
where the parameter $\lambda_0$ is chosen to be positive throughout this work to account for repulsion. Interaction strengths are expressed in terms of the dimensionless mean-field parameter $\Lambda=(N-1)\lambda_0$ with $\lambda_0=\frac{m\,g_{1D}}{\hbar^2n}$ where $n$ is the density and $g_{1D}$ a coupling strength relating the scattering length and the tranverse confinement \cite{Olshanii}. A translation of the dimensionless parameters into real units can be found in Ref. \cite{real_units}.

\subsection{Methodology}\label{methods}
In this section we present a brief description of the linear-response method utilized in order to calculate the many-body excitation spectra of repulsive BECs in one-dimensional lattices. We apply linear-response theory atop the multiconfigurational time-dependent Hartree method for bosons, MCTDHB($M$). The most important ingredient of MCTDHB is the Ansatz of the many-body wavefunction as a superposition of permanents $\{|\vec{n};t\rangle\}$ comprising $M$ single-particle orbitals $\{\phi_q(x,t)|1 \leq q \leq M\}$,
\beq\label{wavefct}
 |\Psi(t)\rangle=\sum_{\vec{n}} C_{\vec{n}}(t)\,|\vec{n};t\rangle,
\eeq
where $\vec{n}=(n_1,\ldots,n_M)^t$ is a vector carrying the individual occupation numbers of the orbitals and $\{C_{\vec{n}}(t)\}$ are the expansion coefficients. The size of the configuration space is thus given by $N_{\text{conf}}=\binom{N+M-1}{N}$. Both the expansion coefficients and the orbitals are time-adaptive and determined by the Dirac-Frenkel time-dependent variational principle. In the case of only one orbital, $M=1$, MCTDHB(1) coincides with the commonly used GP theory where all bosons reside in the same self-consistent single-particle state. 

The linear-response analysis is made atop the static MCHB (multiconfigurational Hartree theory for bosons) ground-state \cite{MCHB} which is obtained by imaginary time-propagation of the MCTDHB equations of motion. Given the ground-state orbitals $\{\phi_q^0\}$ and coefficients $\{C_{\vec{n}}^0\}$, we compute the one- and two-body reduced density matrices $\{\rho_{ij}^0\}$ and $\{\rho_{ijkl}^0\}$. 

Essential for our analysis in section \ref{Results and analysis} is the notion of quantum depletion of the system's ground-state. As a definition, we use the eigenvalues $\{n_k|n_1\geq n_2\geq... \geq n_M\}$ of $\{\rho_{ij}^0\}$, often termed the natural occupation numbers of the eigenvectors which denote the so-called natural orbitals. If only $n_1$ is of order $N$, the system is said to be condensed. Otherwise, it is in general depleted. The depletion $f$ is measured by the sum over all other occupation numbers, 
\beqn\label{eq:Depl}
f= \frac{1}{N}\sum_{k>1}^M n_k.
\eeqn
Explicit values for depletion are given in percent consistently throughout this work.

The linear-response equations are then obtained by linearizing the MCTDHB equations of motion for the ground-state orbitals and coefficients with respect to a small periodic time-dependent perturbation $\delta\hat{h}(x,t)$ to the single-particle Hamiltonian, i.e., $\hat{h}(x)\rightarrow \hat{h}(x)+\delta\hat{h}(x,t)$. This yields the linear-response theory atop MCTDHB($M$), termed LR-MCTDHB($M$). The lengthy but straightforward derivation of the linear-response equations is described in Refs. \citep{LR-MCTDHB1,LR-MCTDHB2}. 

We focus on the resulting eigenvalue equation given by 
\begin{equation}\label{eigenvalue_eq}
	\overline{\mathcal{L}}
	  \begin{pmatrix}\bold{u}^k\\\bold{v}^k\\\bold{C}_u^k\\ \bold{C}_v^k \end{pmatrix}
	 =\omega_k \begin{pmatrix}\bold{u}^k\\\bold{v}^k\\\bold{C}_u^k\\ \bold{C}_v^k \end{pmatrix}
\end{equation}
with the $(2M+N_\text{conf})$-dimensional linear-response matrix $\overline{\mathcal{L}}=\mathcal{M}^{-1/2}\mathcal{P}\mathcal{L}\mathcal{P}\mathcal{M}^{-1/2}$ where $\mathcal{M}$ is a metric containing the reduced one-body density matrix and $\mathcal{P}=\mathbb{1}-\sum_{q=1}^M |\phi_q^0\rangle\langle \phi_q^0|$ is a projection operator onto the subspace orthogonal to the ground-state orbitals. We stress that $\overline{\mathcal{L}}$ is in general not Hermitian. 

The eigenvalue $\omega_k$ denotes the energy of the $k$-th excited state. Due to a special symmtry of $\overline{\mathcal{L}}$, they occur in pairs of one positive and one negative root, i.e.,  $\pm\omega_k$. 

The corresponding eigenvectors $(\bold{u}^k,\bold{v}^k,\bold{C}_u^k,\bold{C}_v^k)^T$ collect the corrections $\bold{u}^k=\{|u_q^k\rangle\}$ and $\bold{v}^k=\{|v_q^k\rangle\}$ to the ground-state orbitals and the corrections $\bold{C}_u^k$ and $\bold{C}_v^k$ to the ground-state coefficients. 

The inner matrix $\mathcal{L}$ is built by four submatrices, 
\begin{equation}\label{LR_mat}
	\mathcal{L}=\begin{pmatrix} \bold{\mathcal{L}}_{oo} & \bold{\mathcal{L}}_{oc} 
					\\ \bold{\mathcal{L}}_{co} & \bold{\mathcal{L}}_{cc}	\end{pmatrix},
\end{equation}
where $\bold{\mathcal{L}}_{oo}$ contains the couplings between the individual orbitals, $\bold{\mathcal{L}}_{oc}$ and $\bold{\mathcal{L}}_{co}$ the couplings between the orbitals and the coefficients, and $\bold{\mathcal{L}}_{cc}$ the couplings between the coefficients solely. The general expressions for these submatrices can be found in Ref. \cite{LR-MCTDHB2}. For the special case of the $\delta$-potential, the $(2M)$-dimensional orbital matrix $\mathcal{L}_{oo}$ is given by 
\begin{equation}\label{entire_Loo}
	\bold{\mathcal{L}}_{oo}=\begin{pmatrix} A & B \\ -B^\ast & -A^\ast  \end{pmatrix}
\end{equation}
with 
\begin{equation}
	A=\{A_{ij}\}, \quad A_{ij}=\rho_{ij}^0\hat{h}+2\lambda_0\sum\limits_{s,l=1}^M\,\rho_{islj}^0\phi_s^{0,\ast} \phi_l^0-\mu_{ij}^0
\end{equation}
and 
\begin{equation}
	B=\{B_{ij}\}, \quad B_{ij}=\sum_{s,l=1}^M\lambda_0\rho_{ijls}^0 \phi_s^0\phi_l^0
\end{equation}
where $\{\mu_{ij}^0\}$ is a Hermitian matrix containing Lagrange mutipliers. 

For $M=1$, $\mathcal{L}$ reduces to the Bogoliubov-de Gennes matrix 
\begin{equation}\label{L_BdG}
	\mathcal{L}_\text{BdG}= \begin{pmatrix} \hat{h}+2\Lambda|\phi^0|^2-\mu^0  & \Lambda \left( \phi^0 \right)^2 \\ -\Lambda \left( \phi^{0,\ast} \right)^2 & -\left(\hat{h}+2\Lambda|\phi^0|^2-\mu^0 \right)^\ast \end{pmatrix} 
\end{equation}
with the chemical potential $\mu^0$. The eigenvalue equation Eq. (\ref{eigenvalue_eq}) therefore yields the particle-conserving BdG equations
\begin{equation}\label{BdG_eq}
	\mathcal{P} \mathcal{L}_\text{BdG} \mathcal{P}\, \begin{pmatrix} u^k\\ v^k \end{pmatrix} = \omega_k \begin{pmatrix} u^k\\ v^k \end{pmatrix}.
\end{equation}
It is important to note that the BdG equations by construction just have access to the single-particle excitations where only one particle at a time is excited out of the GP ground-state. For $\Lambda>0$, we term all excitations computed from Eq. (\ref{BdG_eq}) mean-field excitations. 

In contrast to Eq. (\ref{L_BdG}), the linear-response matrix $\overline{\mathcal{L}}$ from the LR-MCTDHB($M$) theory has a much more complicated structure. Already the orbital matrix $\bold{\mathcal{L}}_{oo}$ from Eq. (\ref{entire_Loo}) is more involved than $\mathcal{L}_\text{BdG}$ because the bosons are permitted to occupy more than just a single orbital. Due to the submatrices $\bold{\mathcal{L}}_{oc}$, $\bold{\mathcal{L}}_{co}$ and $\bold{\mathcal{L}}_{cc}$ in Eq. (\ref{LR_mat}), LR-MCTDHB gives access to additional excitations which are multi-particle in nature, i.e., where more than one particle is excited at a time. Moreover, for non-zero repulsion, even the single-particle excited states within LR-MCTDHB($M>1$) should be more accurate than in the BdG case. This is because it has been shown in Refs. \cite{LR-MCTDHB2,Fetter,Olsen} that given the exact ground-state of the system, a linear-response analysis results in the exact excitation spectrum. The description of the MCHB ground-state however can be systematically improved by increasing the number of orbitals $M$. 

We call all excitations computed from Eq. (\ref{eigenvalue_eq}) many-body excitations since they either (i) describe excitations where more than one particle is excited from the ground-state which can not be obtained from Eq. (\ref{BdG_eq}) directly or (ii) go beyond mean-field theory for $\Lambda>0$, even for single-particle excitations.

We close this section by elaborating briefly on the numerical methods used. The results presented in the next section are obtained by constructing the linear-response matrix $\overline{\mathcal{L}}$ for BECs in one-dimensional lattice potentials with a subsequent diagonalization of $\overline{\mathcal{L}}$. For $M>1$, we perform a partial diagonalization due to the large sizes of the linear-response matrices for the systems considered in this work. To this end, we had to develop a new implementation of LR-MCTDHB which is capable to construct and partly diagonalize such large matrices. Without this, it would not have been possible to obtain converged results for most of the systems in section \ref{Results and analysis}. The implementation uses the {\it Implicitly Restarted Arnoldi Method} (IRAM) \cite{Saad} and its implementation in the ARPACK numerical library \cite{Arpack_HP}. The IRAM is an iterative method that employs the Arnoldi algorithm \cite{Arnoldi} to solve Eq. (\ref{eigenvalue_eq}) with respect to a set of roots which is of special interest. In our case, we consider the lowest-in-energy positive roots, i.e., the low-energy spectrum of many-body excited states. 

For the corresponding BdG matrices a full diagonalization is performed since their sizes are typically much smaller. A more detailed description on the dimensionality of the linear-response matrices used, together with other numerical details, is given Appendix \ref{Appa}.

\section{Results and Analysis}\label{Results and analysis}
In this section, we investigate the spectrum of excited states of interacting BECs in one-dimensional lattice potentials. In section \ref{triple_well} we study spectra in a {\it shallow} and {\it deep} triple well and compare the excitation spectra obtained from the BdG equations with accurate LR-MCTDHB many-body results. In section \ref{larger_systems}, we discuss excitations in larger systems, i.e., with more lattice sites and particles and deduce some general statements on excited states of BECs in one-dimensional lattices. \\

\subsection{Excited states within a triple well}\label{triple_well}

\begin{figure}[t!]
\includegraphics[angle=-90,width=0.55\columnwidth]{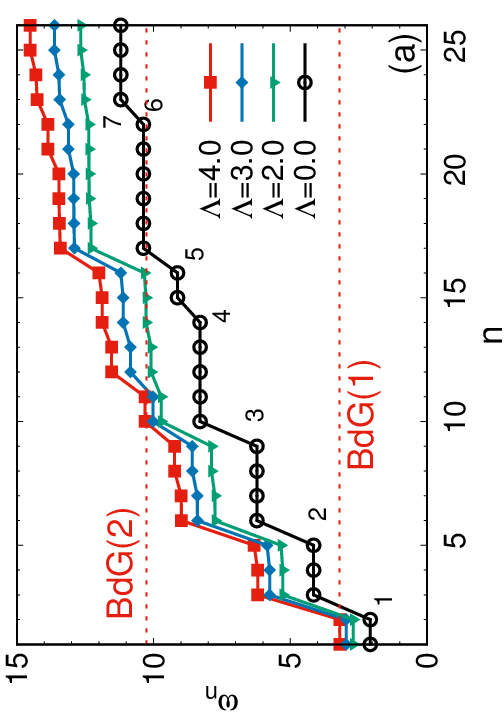}
\includegraphics[angle=-90,width=0.55\columnwidth]{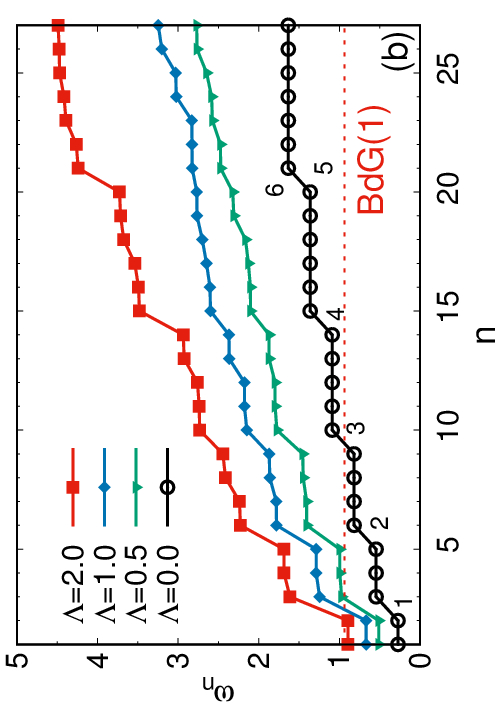}
\caption{(a) Low-energy part of the LR-MCTDHB(7) spectra of excited states for BECs consisting of $N=10$ bosons in a shallow triple well with lattice depth $V_0=5.0$ ($1.01\,E_R$). Excitation energies $\omega_n=E_n-E_0$ are given relative to the ground-state energy $E_0$. Results are shown for different repulsion stengths $\Lambda$. The maximal degree of ground-state depletion is $f=1.1\%$ for $\Lambda=4.0$. For $\Lambda=0$, the states can be separated into distinct levels, labeled in ascending order. The two-fold degenerate states from levels 1 and 5 are single-particle excitations. They can be associated as the states with quasi-momentum $p=\pm 1$ from the first and second single-particle band. All remaining levels solely consist of many-body excitations where more than one particle is excited out of the condensate. By increasing $\Lambda$, the degeneracies between states of the same level are lifted and some levels change the order compared to the non-interacting case. The dotted red lines indicate the first two BdG lines for $\Lambda=4.0$. (b) Same as in (a) but for the deep triple well with depth $V_0=50.0$ ($10.13\,E_R$). The maximal degree of depletion is $f=8.4\%$ for $\Lambda=2.0$, the dotted red line indicates the only BdG line for this repulsion strength in the shown energy range. Notice the different energy scales between the panels. See main text for details. All quantities are dimensionless.}
\label{fig:Triple1}
\end{figure}

\begin{figure}[t!]
\includegraphics[angle=-90,width=0.55\columnwidth]{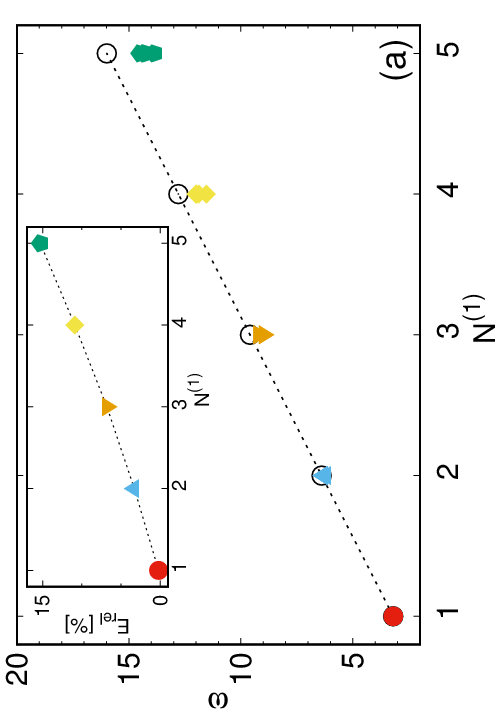}
\includegraphics[angle=-90,width=0.55\columnwidth]{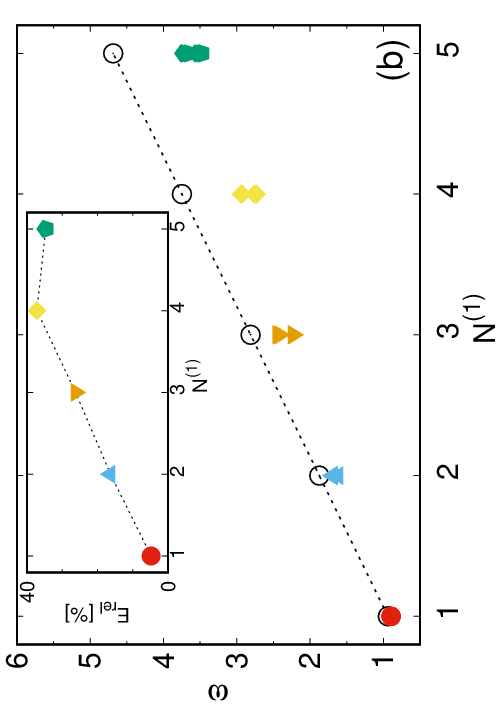}
\caption{(a) Comparison of the first BdG energy BdG(1) and its multiples (open circles) with the LR-MCTDHB(7) many-body results (colored symbols) for a BEC with $N=10$ bosons in the shallow triple well with depth $V_0=5.0$ ($1.01\,E_R$), repulsion strength $\Lambda=4.0$, and depletion $f=1.1\%$ (cp. uppermost curve in Fig. \ref{fig:Triple1}(a)). Shown are the energies $\omega$ of the first five levels where only the $p=\pm1$ states from the first single-particle band are occupied by $N^{(1)}=n_{+1}^{(1)}+n_{-1}^{(1)}$. For the many-body results, the number of points per level reflects the $\left(N^{(1)}+1\right)$-fold degeneracy (note that some points lie atop of each other). The BdG(1) line and its multiples assign too high excitation energies $\omega$ to all levels shown. The deviation grows with $N^{(1)}$. Inset: Evolution of the relative error $E_\text{rel}=\left|\frac{\omega_\text{BdG}\left(N^{(1)}\right)-\omega_\text{MB}\left(N^{(1)}\right)}{\omega_\text{MB}\left(N^{(1)}\right)}\right|$ where $\omega_\text{MB}\left(N^{(1)}\right)$ denotes the LR-MCTDHB(7) energy of the state from the level $N^{(1)}$ with the largest distance to $\omega_{\text{BdG}}\left(N^{(1)}\right)=N^{(1)}\cdot \text{BdG(1)}$. Already for $N^{(1)}=4$, $E_\text{rel}$ exceeds $10\%$. (b) Same as in (a) but for the deep triple well with $V_0=50.0$ ($10.13\,E_R$), $\Lambda=2.0$, and $f=8.4\%$. For this degree of depletion, already BdG(1) itself and the corresponding many-body energy $\omega_\text{MB}(1)$ deviate by a relative error of $E_\text{rel}=4.8\%$ from each other. Notice the different energy scales between the panels. See text for more details. All quantities are dimensionless.}
\label{fig:Triple2}
\end{figure}

To start our analysis, we consider $N=10$ identical bosons confined in a triple well. We distinguish the cases of a shallow and deep triple well with potential depths $V_0=5.0$ ($1.01\,E_R$) and $V_0=50.0$ ($10.13\,E_R$), respectively. Our primary goal is to analyze the general structure of the many-body low-energy spectra and show differences compared to the mean-field spectra obtained by solving the corresponding BdG equations. In a first step, we consider excitations of non-interacting BECs with fully-condensed ground-states and introduce an intuitive protocol for assigning meaningful quantum numbers to the excited states. In a second step, we analyze the effect of interparticle repulsion, and thus ground-state depletion, onto the excited states. We apply LR-MCTDHB(7) to compute the many-body spectra which ensures numerically converged results. A more detailed discussion on the numerical convergence is given in Appendix \ref{Appa}. 

The bottom curve in Fig. \ref{fig:Triple1}(a) shows the energies of the first 26 excited states of the non-interacting BEC in the shallow triple well. The excitation energy values $\omega_n$ are computed relative to the ground-state energy $E_0$, i.e., $\omega_n=E_n-E_0$ where $E_n$ is the energy of the $n$-th excited state. We obtain 7 distinct levels of excitations, each of them composed of a certain number of degenerate states. To guide the eye, these levels are enumerated in ascending order. Within the shown energy range of the figure, we obtain two degenerate pairs of single-particle excitations, given by the doublets of levels 1 and 5. These are single-particle excitations because the corresponding BdG spectrum for the non-interacting case yields only those two pairs. All remaining states can thus only be many-body excitations where more than one particle is excited out of the condensate. In the absence of interparticle repulsion, all single-particle excitations are exact quasi-momentum eigenstates. They appear in bands composed of three states each. The only possible eigenvalues are $p=0$ and $p=\pm 1$, where the states corresponding to the latter are energetically degenerate. 

We therefore identify the doublet from level 1 in Fig. \ref{fig:Triple1}(a) as the $p =\pm1$ eigenstates from the first single-particle band and the doublet from level 5 as the $p =\pm 1$ eigenstates from the second single-particle band. We can now use these states to explain both the energies and the degeneracies of the levels in the spectrum by populating the $p =\pm1$ states with additional particles. To this end, we introduce the notation $\left(n_{+1}^{(1)},n_{-1}^{(1)};n_{+1}^{(2)},n_{-1}^{(2)}\right)$ where the first two entries in brackets denote the occupation of the $p=\pm1$ states from the first single-particle band and the last two entries denote the occupation of the $p=\pm1$ states from the second single-particle band. For example, level 2 is composed of three degenerate states which we identify as the excitations $(2,0;0,0)$ $(0,2;0,0)$ and $(1,1;0,0)$. 

Similar observations can be made for the spectrum of the non-interacting BEC in the deep triple well, given by the bottom curve in Fig. \ref{fig:Triple1}(b). Therein, one obtains only one pair of degenerate mean-field excitations given by the doublet of level 1, representing the $p=\pm 1$ states from the first single-particle band. One thus needs only this level to explain the positions of the many-body levels in the figure. The $p=\pm 1$ states from the second single-particle band are significantly higher in energy ($\omega \approx 24.1$, not shown).

Let us now consider repulsion between the bosons and study how this affects the excitation spectra. In contrast to the fully-condensed ground-states of the non-interacting systems, there are now different degrees of ground-state depletion depending on the repulsion strength $\Lambda$. For the shallow lattice, we obtain $f=0.4\%$ for $\Lambda=2.0$, $f=0.73\%$ for $\Lambda=3.0$, and $f=1.1\%$ for $\Lambda=4.0$. We thus claim to be in the limit of weak depletion for all values of $\Lambda$ considered for the shallow lattice. For the deep lattice, the ground-states in general show stronger depletion, given by $f=2.3\%$ for $\Lambda=0.5$, $f=4.7\%$ for $\Lambda=1.0$, and $f=8.4\%$ for $\Lambda=2.0$. 

As a general observation, one can see from the upper three curves in Fig. \ref{fig:Triple1}(a) and (b) that the low-energy spectra of the shallow and deep lattices are growing monotoneously in energy when the repulsion is increased, i.e., the distance $\omega_n$ to the ground-state becomes larger for all excited states shown. Moreover, the order of levels in the shallow triple well changes for growing values of $\Lambda$. For example, the third many-body level which solely consists of multi-particle excitations (states $n=12$ to $16$) crosses the second single-particle level (states $n=10$ and $11$). Also levels 6 and 7 switch the order compared to the non-interacting case. 

In the following, we concentrate on many-body effects of the excitation spectra for the cases $\Lambda=4.0$ in the shallow triple well and $\Lambda=2.0$ in the deep triple well only. Comparing the mean-field and many-body results, one can see that the low-energy spectrum consists of a large number of many-body excited states. For the $\Lambda=4.0$ case in the shallow triple well, two BdG lines for the energy range considered are obtained (indicated by the dotted horizontal red lines in Fig. \ref{fig:Triple1}(a)). For the $\Lambda=2.0$ case in the deep triple well we obtain one BdG line (dotted red line in Fig. \ref{fig:Triple1}(b)). As for $\Lambda=0$, we identify those states as the $p=\pm 1$ states from the first and second single-particle bands. Whereas this identification was exact in the non-interacting systems, this is not the case any more for $\Lambda>0$ since now the states are dressed by the interaction potential. However, we keep the notion $\left(n_{+1}^{(1)},n_{-1}^{(1)};n_{+1}^{(2)},n_{-1}^{(2)}\right)$ to label states from the low-energy spectrum because the upper curves in Fig. \ref{fig:Triple1}(a) and (b) show that the excitations can still be classified according to this level structure.

The first many-body effect one can observe is the splitting of states from the same level in the presence of interaction. For example, the state $n=5$ in the upper curve of Fig. \ref{fig:Triple1}(a) is slightly higher in energy ($\omega_5=6.32$) than the states $n=3,4$ with $\omega_{3,4}=6.19$. Due to symmetry reasons, we identify the two degenerate states as $(2,0;0,0)$ and $(0,2;0,0)$, whereas the non-degenrate state is $(1,1;0,0)$. The splitting grows with $\Lambda$. It is important to note that in both the non-interacting and mean-field cases, all states from the same level are degenerate, and a distinction like above is not possible. Thus, the splittings in the many-body spectrum allow at least partly to identify individual states of the same many-body level. More generally, we observe that states where the modulus between the total quasi-momentum $P$ given by 
\begin{equation}\label{Tot_Quasi}
	P=n_{+1}^{(1)}+n_{+1}^{(2)}-n_{-1}^{(1)}-n_{-1}^{(2)}
\end{equation}
and the number of sites $L$ equals zero, i.e., $\text{mod}(P,L)=0$, are not degenerate. We will call these states the zero quasi-momentum modes, or simply ZQMs, in the following. A more detailed discussion on the ZQMs in the shallow triple well is given in Appendix \ref{AppB}. States from the same level with equal $|\text{mod}(P,L)|\neq0$ remain two-fold degenerate. 

As a second many-body effect, we discuss the numerical deviations of the BdG excitation energies from the ones obtained from LR-MCTDHB(7). As mentioned in Sec. \ref{sec:Intro}, the BdG energies give the exact values for the single-particle excitations in the infinite-particle limit, i.e., for $N\rightarrow \infty$ and $\lambda_0\rightarrow 0$ such that $\Lambda=(N-1)\cdot \lambda_0=\text{const.}$ Again, the energies for multi-particle excitations in this limit are exactly given by multiples and sums of the BdG energies.

Fig. \ref{fig:Triple2} shows the deviations between the first BdG line, BdG(1), and its first few multiples from the corresponding exact many-body results for (a) the shallow triple well with $\Lambda=4.0$ and (b) the deep triple well with $\Lambda=2.0$. For both systems, we already find clear deviations for the first two mean-field-like excitations, $n=1$ and $2$. The exact many-body energies for both systems are smaller than the corresponding BdG energies. The reason is that due to the depleted ground-state in both cases, there are now slightly less particles in the condensed mode, and thus the effective repulsion between the bosons in that mode is weakened. Many-body linear-response atop the depleted ground-state accounts for this lower repulsion and, in comparison with the mean-field result, yields a slightly lower excitation energy. 

Another quantitative measure for the deviations is given by the relative error $E_\text{rel}$ (see details on the computation of $E_\text{rel}$ in the caption of Fig. \ref{fig:Triple2}). Whereas it is still moderate for the shallow triple well, given by $E_\text{rel}=0.3\%$ for the first and $E_\text{rel}=0.5\%$ for the second BdG line (not shown), it becomes significantly larger in the deep lattice where we obtain $E_\text{rel}=4.8\%$ for the first BdG line, cp. the insets of Fig. \ref{fig:Triple2}(a) and (b). In other words, already for this first level of excited states the energies of the mean-field and many-body spectra deviate substantially from each other. 


With respect to the multi-particle excited states, the deviations become even larger. If the mean-field approximation using multiples of the first BdG line (open circles in Fig. \ref{fig:Triple2}) would be exact, all energies from the many-body spectrum (colored symbols) would follow the straight line and lie atop the BdG(1) multiples. However, in both cases (a) and (b), the mean-field approximation tends to overestimate the exact energies of the individual levels. The increase in the deviation between multi-particle excitations and their corresponding BdG multiples are in line with the above analysis of lowered effective repulsion between the bosons remaining in the condensed mode of the depleted ground-state. Moreover, it assigns by construction the same energy to all excitations from the same level, which contradicts to the previously made observation of splittings. For the stronger depleted condensate in (b), the descrepancies are substantial and the relative error given in the inset exceeds $10\%$ already for $N^{(1)}=2$. But also for the weakly depleted condensate in (a) the relative error grows quickly, given by $E_\text{rel}=10.9\%$ and $15.3\%$ for the levels $N^{(1)}=4$ and $5$, respectively.

\subsection{Larger systems}\label{larger_systems}

After having studied the many-body excitation spectra of repulsive BECs in a triple well, we now investigate excitations in larger systems, i.e., with more lattice sites and particles. Treating below more particles and sites aims at solidifying the above found many-body effects and announcing better computational capabilities of many-body excitations spectra than reported previously in \cite{LR-MCTDHB1,Theisen}.

\begin{figure}[t!]
\includegraphics[angle=-90,width=0.55\columnwidth]{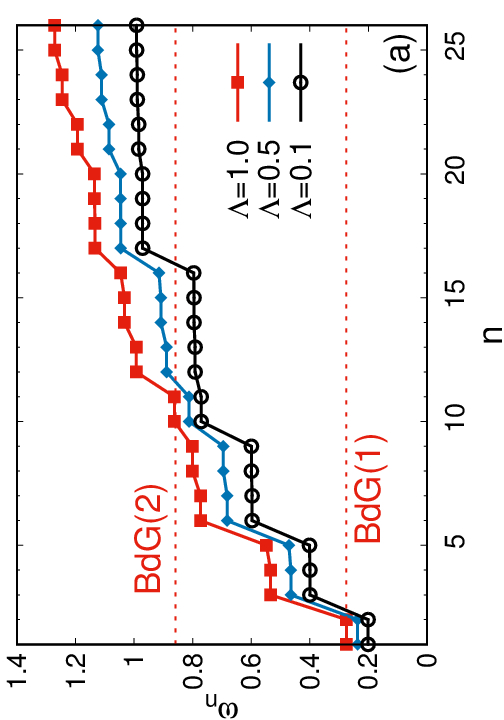}
\includegraphics[angle=-90,width=0.55\columnwidth]{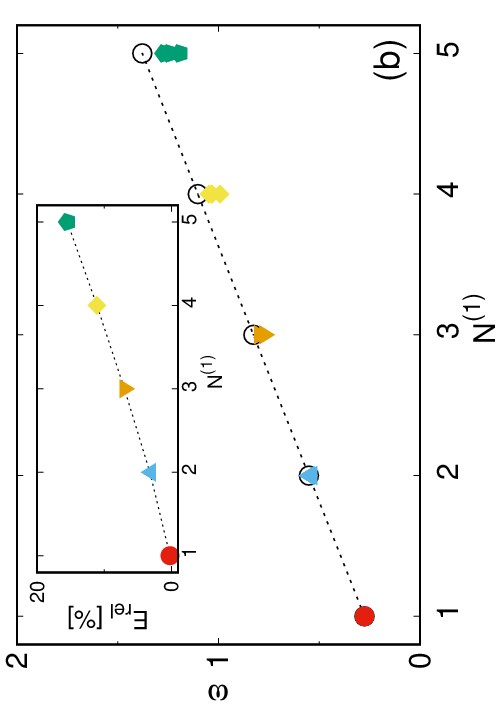}
\caption{(a) Same as in Fig. \ref{fig:Triple1}(a) but for a lattice with 10 sites. The maximal degree of ground-state depletion is $f=0.84\%$ for $\Lambda=1.0$. The states $n=1,2$ and $n=10,11$ are single-particle excitations, all remaining states of the many-body spectra are multi-particle excitations where more than one boson at a time is excited from the ground-state. With increasing $\Lambda$, several degeneracies of states from the same level are lifted compared to the non-interacting case. Dotted red lines indicate the two BdG mean-field energies for $\Lambda=1.0$ in the shown energy range. (b) Same as in Fig. \ref{fig:Triple2}(a) but for a lattice with 10 sites and repulsion strength $\Lambda=1.0$. The BdG(1) line and its multiples assign too high excitation energies $\omega$ to all levels shown. The deviation grows with $N^{(1)}$. Inset: Evolution of the relative error $E_\text{rel}$. See text for more details. All quantities are dimensionless.}
\label{fig:large1}
\end{figure}

Fig. \ref{fig:large1}(a) shows many-body excitation spectra for BECs with $N=10$ bosons in a shallow lattice ($V_0=5.0=1.01 E_R$) with 10 sites. The ground-state depletion for the different repulsion strengths considered are given by $f=0.02\%$ for $\Lambda=0.1$, $f=0.29\%$ for $\Lambda=0.5$, and $f=0.84\%$ for $\Lambda=1.0$. We are thus dealing with excited states in a similar regime of weak depletion as we did for the shallow triple well in the previous section. The many-body results are numerically converged for LR-MCTDHB(7) (see discussion in Appendix \ref{Appa}).

As for the triple well cases, the low-energy spectrum contains a large number of many-body excitations, and the spectrum grows monotoneously in energy with growing repulsion. The excited states $n=1,2$ and $n=10,11$ are close to the first two BdG lines. We identify these single-particle states as quasi-momentum eigenstates, dressed by the interaction potential. However, in the larger lattice the spectrum of quasi-momentum eigenstates is denser, allowing for bands of 10 states with momenta $p\in [-5,4]$. Therefore, the states $n=1,2$ correspond to the dressed $p=\pm 1$ eigenstates and the states $n=10,11$ correspond to the dressed $p=\pm 2$ eigenstates from the first single-particle band. All remaining states denote many-body excitations with more than one particle being excited at a time. Similar to the triple well cases, the splitting of states with different momenta becomes larger with increased repulsion. 

Fig. \ref{fig:large1}(b) presents the comparison between the mean-field approximation, i.e., BdG(1) line and its multiples, with the many-body results for the first five levels, build by occupying the $p\pm 1$ states from the first single-particle band for the $\Lambda=1.0$ spectrum from panel (a). As for the triple well examples, the multiples of BdG(1) assign too high excitation energies to the individual levels and the deviation grows with $N^{(1)}$. It is very interesting to see from the inset that the evolution of the relative error yields very similar values as the ones from the $\Lambda=4.0$ case in the shallow triple well where the degree of the ground-state depletion was of the same order (cp. inset of Fig. \ref{fig:Triple2}(a)). However, we emphasize that in the lattice with 10 sites the deviations appear already for weaker repulsion. The physical origin lies in the denser spectrum of quasi-momentum eigenstates compared to the triple well, such that already for weak repulsion, several quasi-momentum eigenstates are occupied. 

So far we considered systems with $N=10$ bosons. We now increase the number of particles. The motivation is to identify convincing differences between the mean-field and many-body spectra for a system which is closer to the infinite-particle limit of the trapped condensate. 

\begin{figure}[t!]
\includegraphics[angle=-90,width=0.55\columnwidth]{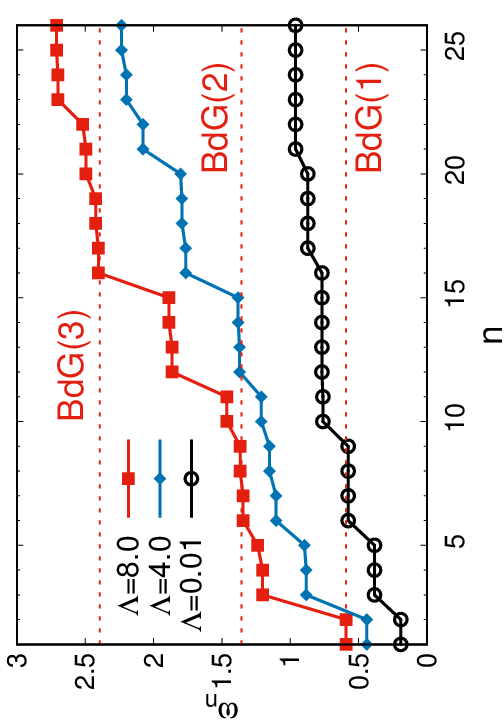}
\caption{Many-body excited states for a BEC with $N=100$ bosons in a lattice with 10 sites and depth $V_0=5.0$ ($1.01\,E_R$). Results are shown for different repulsion strength $\Lambda$ and have been calculated with LR-MCTDHB(3). The maximal depletion is $f= 1.3\%$ for $\Lambda=8.0$. Dotted red lines indicate the BdG mean-field energies for $\Lambda=8.0$ in the shown energy range. The many-body spectra show similar effects like for the smaller systems with only ten particles, e.g., the splitting of states from the same level. See text for more details. All quantities are dimensionless.}
\label{fig:large2}
\end{figure}

Fig. \ref{fig:large2} shows the excitation spectra for $N=100$ bosons in a shallow lattice with 10 sites for several interaction strengths. We calculated the many-body energies with LR-MCTDHB(3). In contrast to all results shown so far for $N=10$ particles, we do not claim to present fully-converged energies in this case. One probably needs to include $M=5$ single-particle orbitals or more into the linear-response analysis to obtain fully converged excitation energies. Since the size of the corresponding coefficients' subspace is larger than $9\cdot 10^6$ and, thus, the numerical effort to compute the low-energy spectrum for $M \geq 5$ exceeds the scope of this work. Nonetheless, the results are sufficient to show that clear deviations from BdG theory are obtained because the quality of the ground-state is improved. In particular, adding more orbitals to the description of the ground-state cannot remove the many-body features in the excitation spectra.  

The degrees of ground-state depletion is given by $f=10^{-5}\%$ for $\Lambda=0.01$, $f=0.36\%$ for $\Lambda=4.0$, and $f=1.3\%$ for $\Lambda=8.0$. We are therefore within the same regime of depletion as for the cases of $N=10$ bosons in shallow lattices discussed above. 

Firstly, we again see in Fig. \ref{fig:large2} that a large number of many-body excitations appear in the low-energy part of the spectrum. Secondly, with increasing repulsion, several splittings of excitations from the same level is observed, showing that this many-body effect also appears in systems which are already closer to the infinite-particle limit than the examples with $N=10$ bosons. Moreover, they occur within the same regime of depletion which is around $f\approx 1\%$. With respect to the repulsion strength, however, depletion sets in for higher values of $\Lambda$ because the closer one gets to the infinite-particle limit, the more condensed is the system's ground-state for a fixed interaction strength and thus one needs to increase the repulsion to obtain stronger ground-state depletion.

\section{Summary and Conclusions}\label{Conclusions}

In this work, we have investigated excitation spectra of repulsive BECs in one-dimensional lattice potentials with periodic boundary conditions. Our study dealt with the low-energy part of the spectra on the mean-field and many-body level, computed from the BdG equations and LR-MCTDHB, respectively. For all systems considered, we saw that the LR-MCTDHB low-energy spectra consist of a large number of many-body excitations and many-body properties which the BdG equations can not access by construction. 

We presented numerically accurate results for excitation spectra of BECs with $N=10$ bosons in shallow and deep triple wells. In both cases, many-body effects on the excitation spectra set in as soon as there is ground-state depletion of the order of about $1\%$. Those effects are mainly (i) the splitting between several many-body excitations from the same level which are degenerate in the corresponding mean-field spectra, (ii) the numerical deviations between the BdG lines and the corresponding many-body results from LR-MCTDHB when the ground-state is depleted, and as a consequence of this, we found (iii) that the multiples of the BdG lines do not accurately account for the excitation energies of excited states where more than one boson at a time is excited. The mean-field approximation of taking multiples of the BdG energies quickly becomes inaccurate, i.e., even for the lowest levels one can see substantial differences compared to the many-body results. This observation was made already for weakly-depleted condensates with $f\approx1\%$. We conclude that one is clearly in need of an accurate many-body description for the excited states, even in the regime of weak depletion. 

We extended our study to systems with additional lattice sites and particles. For $N=10$ bosons in a shallow lattice with 10 sites, we found that the many-body effects described above set in already for weaker repulsion than for the triple-well examples. This is because the spectrum of low-energy quasi-momentum eigenstates is denser for the larger lattice. From this we deduce that, in general, the excitation spectra of condensates with a finite number of particles in larger lattices are even more sensitive to the interaction strengths, i.e., many-body effects set in at weaker values of $\Lambda$ than in smaller lattices.

For a larger system with $N=100$ bosons, i.e., closer to the infinite-particle limit, we obtain qualitatively similar excitation spectra, containing clear many-body effects. However, the repulsion strength needs to be larger in order to achieve the same degree of ground-state depletion. We have again demonstrated that once about $1\%$ is depleted, many-body effects can no longer be neglected.


\section*{Acknowledgments}

Computation time on the Cray  XC40 cluster Hazel Hen 
at the High Performance Computing Center Stuttgart (HLRS) is achknowledged. Financial support by the IMPRS-QD (International Max Plack Research School for Quantum Dynamics), the Landesgraduiertenf{\"o}rderung Baden-W{\"u}rttemberg, and the Minerva Foundation are gratefully acknowledged. O. E. A. ackowledges funding by the Israel Science Foundation (Grant No. 600/15).

\appendix

\section{Details of the computations and numerical convergence}\label{Appa}

In this section, we present additional details on the numerical computations and report on the numerical convergence for the excitation spectra for all systems with $N=10$ bosons discussed in the main text. 

In general, our computations have been carried out on a grid with 16 or more discrete-variable-representation (DVR) grid points per lattice site. For all simulations done, it turned out that this is sufficient since increasing the number of grid points per site did not change the obtained excitation energies. The box sizes used are $[-1.5,1.5)$ for the triple wells and $[-5,5)$ for the lattice with 10 sites. 

To obtain the MCHB ground-states from propagating the MCTDHB equations of motion in imaginary time, we use the MCTDHB implementation in the software package \citep{package}, also available in the recently developed MCTDHB-Laboratory package \cite{MCTDHB-lab}. 

\begin{figure}[t!]
\includegraphics[angle=-90,width=0.55\columnwidth]{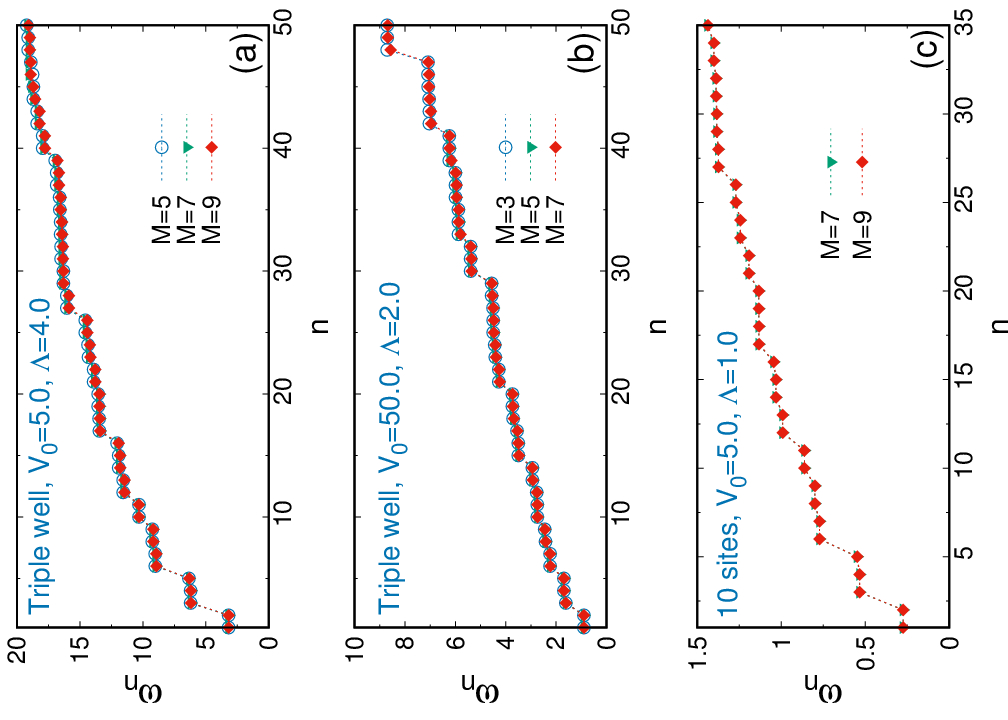}
\caption{(a) LR-MCTDHB spectra for a BEC with $N=10$ in a shallow triple well with $V_0=5.0$ ($1.01\,E_R$) with repulsion strength $\Lambda=4.0$ for different numbers of single-particle orbitals $M$. The spectra for $M=7$ and $M=9$ lie atop of each other, indicating numerical convergence for $M=7$ orbitals. (b) Same as in (a) but in a deep triple well with $V_0=50.0$ ($10.13\,E_R$) with repulsion strength $\Lambda=2.0$. Already the LR-MCTDHB(3) spectrum lies atop of the spectrum for $M=7$, with only very little exceptions for the higher lying states. (c) Same as in (a) but for 10 lattice sites and repulsion strength $\Lambda=1.0$. Notice the different energy scales between the panels. See text for more details. All quantities are dimensionless.}
\label{fig:Conv}
\end{figure}

The linear-response matrices of the systems studied in this work are of the dimensionality
\begin{equation}
 	N_\text{dim}=2M\cdot N_\text{DVR}  +2N_\text{conf}
\end{equation}
where $N_\text{DVR}$ is the total number of DVR grid points used. Typically, the BdG matrices considered are small in size, i.e., $N_\text{dim}(\mathcal{L}_\text{BdG})<1000$. We therefore perform a full diagonalization of these matrices. However, the linear-response matrices for the many-body cases are usually much larger. For example, the triple-well examples in the main text are carried out on a grid with 32 DVR points per site, yielding $N_\text{dim}=17360$ for $M=7$ and $N_\text{dim}=89244$ for $M=9$ (see below and Fig. \ref{fig:Conv}). In order to deal also with such large matrices, we extended our LR-MCTDHB implementation from Refs. \citep{LR-MCTDHB1,LR-MCTDHB2,LR-MCTDHB3}, both with respect to the construction and (partly) diagonalization of $\overline{\mathcal{L}}$, which is planned to be reported and published elsewhere. 

Next we discuss the numerical convergence with respect to the number $M$ of single-particle orbitals used. Fig. \ref{fig:Conv}(a) shows the low-energy spectrum for the same system as in the upper curve of Fig. \ref{fig:Triple1}(a) for different numbers of orbitals $M$. Already for $M=5$ orbitals, the results obtained agree very nicely with the more involved computations done with $M=7$ and $M=9$ orbitals. Only slight deviations can be observed. Most importantly, the spectra for $M=7$ and $M=9$ coincide, indicating that the usage of $M=7$ orbitals for the quantitative analysis in the main text is justified. 

The same conclusion can be made for the example in Fig. \ref{fig:Conv}(c) where numerical convergence with $M=7$ orbitals is proven for the system of the upper curve of Fig. \ref{fig:large1}(a). Including another two orbitals, i.e., $M=9$ in total, shows no substantial differences with respect to the resulting energies of excited states. 

Fig. \ref{fig:Conv}(b) shows a rather exceptional situation for the numerical convergence of the system from the upper curve of Fig. \ref{fig:Triple1}(b), i.e., $N=10$ bosons in the deep triple well with interaction strength $\Lambda=2.0$. We emphasize again that the corresponding ground-state is depleted with $f=8.4\%$. Although the depletion in this case is clearly higher than for the two systems from Fig. \ref{fig:Conv}(a)+(c), already for $M=3$ one obtains numerically accurate results for the excitation energies. It has not been the case for the BECs in the shallow lattices. The reason might be that the BdG lines in the low-energy part of the spectrum are in this situation very far apart from each other ($\text{BdG(1)}=9.3$ and $\text{BdG(2)}=23.8$). Due to this, the MCHB(3) ground-state is most likely composed of three single-particle orbitals that are very close to the three states of the first single-particle band from which all many-body excited states of the low-energy spectrum are build. Since for the BECs in the shallow lattices the first two mean-field levels are not that much separated from each other, one needs at least $M=5$ orbitals in order to obtain reliable results. 

Most importantly, in all cases considered, the many-body effects atop BdG persist as the quality of the ground-state improves.

\section{Zero quasi-momentum modes}\label{AppB}

\begin{figure}[t!]
\includegraphics[angle=-90,width=0.55\columnwidth]{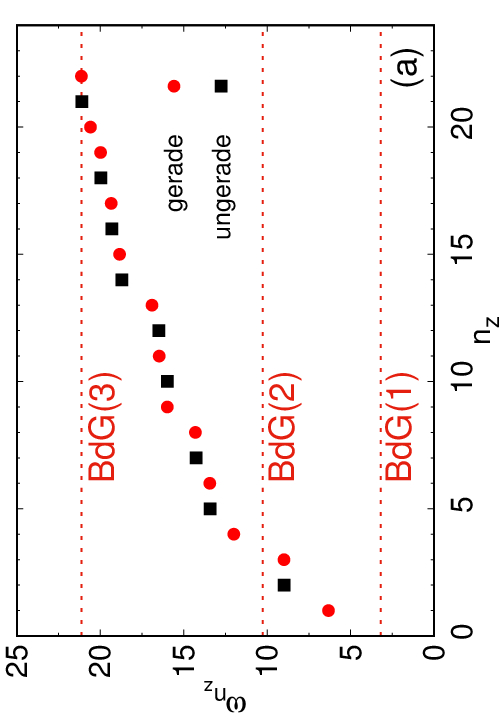}
\includegraphics[angle=-90,width=0.55\columnwidth]{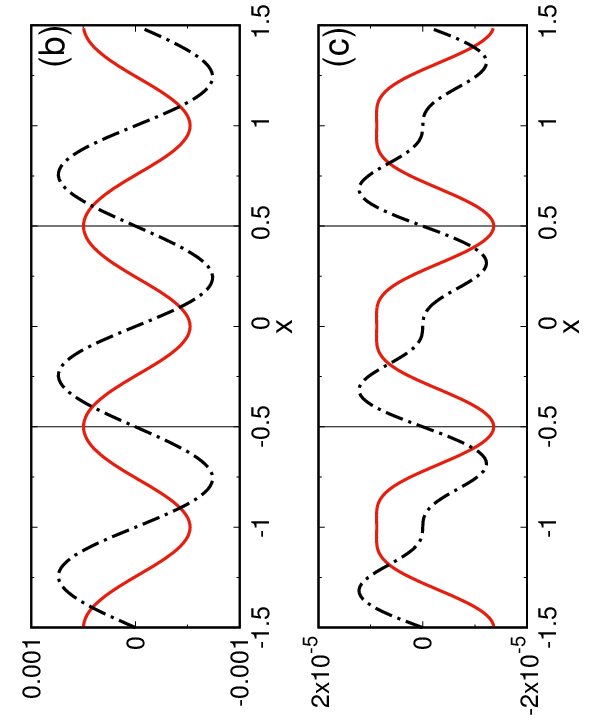}
\caption{(a) Energies of the zero quasi-momentum modes (ZQMs) for the system with $N=10$ repulsive bosons with $\Lambda=4.0$ in the shallow triple well with $V_0=5.0$ ($1.01\,E_R$). Calculations are carried out with LR-MCTDHB(7). Shown are all ZQMs up to the third BdG mean-field excited state $\text{BdG}(3)$ which indicates the top of the second single-particle band. All ZQMs can be characterized by either gerade (red dots) or ungerade (black squares) symmetry. All states shown are many-body excitations and thus not included in the corresponding mean-field spectrum. Note the different index $n_z$ which enumerates only the ZQMs compared to the index $n$ in the figures of the main text which enumerates all excitations. (b) Real part of the density responses for the states $(3,0;0,0)$ and $(0,3;0,0)$. (c) Real part of the density responses for the states $(4,1;0,0)$ and $(1,4;0,0)$. Vertical lines separate the lattice sites. Notice the different scales between panels (b) and (c). See text for more details. All quantities are dimensionless.}
\label{fig:zeroes}
\end{figure}

In this section, we elaborate on the zero quasi-momentum modes (ZQMs) of the system with $N=10$ bosons in the shallow triple well with interaction strength $\Lambda=4.0$. As a reminder, we define the ZQMs as the states where $\text{mod}(P,L)=0$, i.e., where the modulus of the total quasi-momentum $P$ given in Eq. (\ref{Tot_Quasi}) and the number of sites $L$ equals zero.

In order to analyze the shape of the ZQMs, we calculate the response densities
\begin{equation}\label{response_dens}
	\Delta\rho^k(x)=\Delta\rho_o^k(x)+\Delta\rho_c^k(x)
\end{equation}
which have been introduced in \cite{LR-MCTDHB1}. In Eq. (\ref{response_dens}), the quantities $\Delta\rho_o^k(x)$ and $\Delta\rho_c^k(x)$ denote the orbitals' and coefficients' contribution to the total response density of the excited state $k$, respectively. For general stationary orbitals $\left\{\phi_i^0(x)\right\}$ and reduced one-body density matrix $\pmb{\rho}^0=\{\rho_{ij}^0\}$ these contributions are given by
\begin{align}
		\Delta\rho_o^k&=\left(\Phi^{0}\right)^\dagger(\pmb{\rho}^0)^{1/2}\left( \bold{u}^k+\bold{v}^{k,\ast}\right)  \nonumber \\
		&+\left( \bold{u}^{k}+\bold{v}^{k,\ast}\right)^\dagger(\pmb{\rho}^{0,\ast})^{-1/2}\pmb{\rho}^{0}\, \Phi^0 
\end{align}
and
\begin{equation}
		\Delta\rho_c^k=\sum_{i,j=1}^M \, \phi_i^{0,\ast}\phi_j^{0} \left( \langle \bold{C}^0 |\hat{a}_i^\dagger \hat{a}_j| \bold{C}_u^k \rangle+\langle \bold{C}_v^{k,\ast} |\hat{a}_i^\dagger \hat{a}_j| \bold{C}^0 \rangle \right),
\end{equation}
where the vector $\Phi^{0}=\left(\phi_1^0,...,\phi_M^0\right)^t$ collects the stationary orbitals. All position arguments in the latter two equations have been omitted for the sake of simplicity. The response densities should not be confused with the actual densities of the excited states. Whereas the latter are normalized to unity and real-valued, the response densities from Eq. (\ref{response_dens}) are not normalized and in general complex. Their intensities signify the strength of the excitations' contributions to the response wavefunction \cite{LR-MCTDHB1,LR-MCTDHB2}. 

Fig. \ref{fig:zeroes}(a) shows the energies of the ZQMs up to the top of the second single-particle band, given by the third BdG line, BdG(3). We obtain in total 22 ZQMs up to this energy where especially between the BdG(2) and BdG(3) lines the density of ZQMs is high. We stress that all of these states are many-body excitations, and therefore not included in the corresponding mean-field spectrum. The response densities indicate that the ZQMs can be characterized by their spatial symmetry which appears to be either gerade or ungerade with respect to the individual lattice sites. We observe that for all pairs of ZQMs from the same many-body level and the same magnitude $|P|$, one state of the pair is gerade and the other one is ungerade. It explains why these ZQMs are non-degenerate. As an example, Fig. \ref{fig:zeroes}(b) and (c) show the real parts of the response densities for two of such ZQM pairs. In particular, several other ZQMs from the spectrum in panel (a) have the same shape as the response density of the gerade excitation in panel (b). One can assume that these states can be excited quite easily in an experiment. For example, a slight driving of the depth $V_0(t)=V_0+\Delta_{V_0}\sin(\omega t)$ with driving frequency $\omega$ and amplitude $\Delta_{V_0}$ should populate those excitations, especially when $\omega$ is in the vicinity of the energy $\omega_{n_z}$ of such an excited state. \\

\section{Benchmark to the one-dimensional harmonic interaction model}\label{AppC}

To benchmark our numerical implementation of LR-MCTDHB, we use the one-\break\hfill dimensional harmonic interaction model (1D-HIM) and compare the numerical results with the exact analytic excitation energies. In the 1D-HIM, both the trapping potential $\hat{V}(x)=\frac{1}{2}\Omega x^2$ and the two-body interaction potential $\hat{W}(|x_i-x_j|)=\lambda_0|x_i-x_j|^2$ are of harmonic type. The analytic excitation energies can be found in, e.g., Ref. \cite{Cohen}. \\
Table \ref{table2} shows results for a BEC with $N=10$ bosons in a trap with trapping frequency $\Omega=1.0$ and interaction strength $\lambda_0=0.13$ which yields a comparable degree of depletion ($f=0.94\%$) as for the systems in the main text. LR-MCTDHB(1)$\equiv$BdG only gives a converged result for the first center-of-mass excitation $\omega_1$. All other excitations that are accessible within BdG show clear deviations from the analytic values. By adding additional orbitals to the description of the ground-state, all missing lines are obtained, and the accuracy of all excitations is clearly improved. For $M=6$ orbitals, the agreement with the analytic values becomes highly accurate for the first ten excitations.


\begin{table}[t!]
\begin{tabular*}{0.65\textwidth}{p{1.3cm} p{1.8cm} p{1.8cm} p{1.8cm} c}

 \hline\hline
 	&	$M=1$	&	$M=4	$	&	$M=6$	& 	Exact  analytical	 \\ 
 \hline  
$E_0$	&	9.\underline{137833}		&	9.03815\underline{1}	&	9.038150		&	9.038150		 \\ 	
$\omega_1$	&	1.000000		&	1.000000	&	1.000000		&	1.000000		 \\ 
$\omega_2$	&	n/a	&	2.000\underline{222}	&	2.000000		&	2.000000		 \\ 
$\omega_3$	&	n/a	&	3.000\underline{432}	&	3.000000	    &	3.000000		\\ 
$\omega_4$	&	3.\underline{655133}	&	3.7947\underline{52}	&	3.794733 	&	3.794733	\\ 
$\omega_5$	&	n/a	&	4.0\underline{11839}	&	4.0000\underline{12}	 &	4.000000 \\ 
$\omega_6$	&	n/a	&	4.794\underline{870}	& 4.794733 &	 4.794733	\\ 
$\omega_7$	&	n/a	&	5.0\underline{22646}	&	5.0000\underline{20}	&	5.000000 \\ 
$\omega_8$	&	5.\underline{482700} 	&	5.6921\underline{43}	& 5.692100 &	 5.692100	\\ 
$\omega_9$	&	n/a	&	5.79\underline{7912}	& 5.79473\underline{7} &	 5.794733	\\ 
$\omega_{10}$	&	n/a	&	6.\underline{142955}	& 6.000\underline{624} &	 6.000000	\\ \hline\hline
\end{tabular*}
\caption{Benchmark of the numerical method for $N=10$ bosons in the one-dimensional harmonic interaction model. Shown are the ground-state energy $E_0$ and the energies $\omega_i$ of the first ten excitations. The trapping frequency is $\Omega=1.0$ and the interaction strength is $\lambda_0=0.13$, yielding ground-state depletion of $f=0.94\%$ which is in the same regime as for the systems discussed in the main text. Underlined digits denote deviations from the exact values. See text for more details. All quantities are dimensionless.}
\label{table2}
\end{table}

\newpage



\begin{thebibliography}{99}
\bibitem{real_units}	To convert the dimensionless parameters to real units, consider for instance a generating wave of the lattice with wavelength $\lambda=700$ nm which yields a spacing of $l=0.35\mu$m between two sites. The recoil frequency $\omega_R=E_R/\hbar$ is thus given by 29.4 kHz. The interaction between the bosons can be regulated by changing the scattering length $a_s$ with a Feshbach resonance. For a BEC of $N=10$ interacting $^{87}$Rb atoms with mass $86.909187$ u (NIST), the frequnency of the transverse confinement $\omega_\perp$ and $a_s$ can be chosen as follows. For the triple well and $\Lambda=4.0$: $\omega_\perp=40$ kHz and $a_s=27$ nm. For ten sites and $\Lambda=1.0$: $\omega_\perp=20$ kHz and $a_s=5$ nm.
\end{thebibliography}
\end{document}